\documentclass[10pt,twocolumn,letterpaper]{article}
\usepackage[pagenumbers]{iccv} 
\usepackage[accsupp]{axessibility}  
\usepackage{times}
\usepackage{epsfig}
\usepackage{graphicx}
\usepackage{amsmath}
\usepackage{amssymb}
\usepackage{comment}
\usepackage{tabularx}
\usepackage{dsfont}

\definecolor{iccvblue}{rgb}{0.21,0.49,0.74}
\usepackage[pagebackref,breaklinks,colorlinks,allcolors=iccvblue]{hyperref}

\def\paperID{16007} 
\def\confName{ICCV}
\def\confYear{2025}

\title{Inverse 3D Microscopy Rendering for Cell Shape Inference with Active Mesh} 

\author{Sacha Ichbiah \qquad Anshuman Sinha \qquad Fabrice Delbary \qquad Hervé Turlier\\
\textit{Collège de France, CNRS, Inserm, PSL University}, Paris, France \\
{\tt\small herve.turlier@cnrs.fr}}

\begin{document}
\maketitle

\begin{abstract}
   Traditional methods for biological shape inference, such as deep learning (DL) and active contour models, face important limitations in 3D. DL approaches require large annotated datasets, which are often impractical to obtain, while active contour methods depend on carefully tuned heuristics for intensity attraction and shape regularization. We introduce deltaMic, a novel differentiable 3D renderer for fluorescence microscopy that formulates shape inference as an inverse problem. By leveraging differentiable convolutions, deltaMic simulates the image formation process, integrating a parameterized point spread function (PSF) with a triangle mesh-based representation of biological structures. Unlike DL- or contour-based segmentation, deltaMic directly optimizes both shape and optical parameters to align synthetic and real microscopy images, removing the need for large datasets or sample-specific fine-tuning. To ensure scalability, we implement a GPU-accelerated Fourier transform for triangle meshes along with narrow-band spectral filtering. We show that deltaMic accurately reconstructs cell geometries from both synthetic and diverse experimental 3D microscopy data, while remaining robust to noise and initialization. This establishes a new physics-informed framework for biophysical image analysis and inverse modeling.
\end{abstract}

\section{Introduction}
\begin{figure*}[h]
    \centering
    \includegraphics[width=0.95\linewidth]{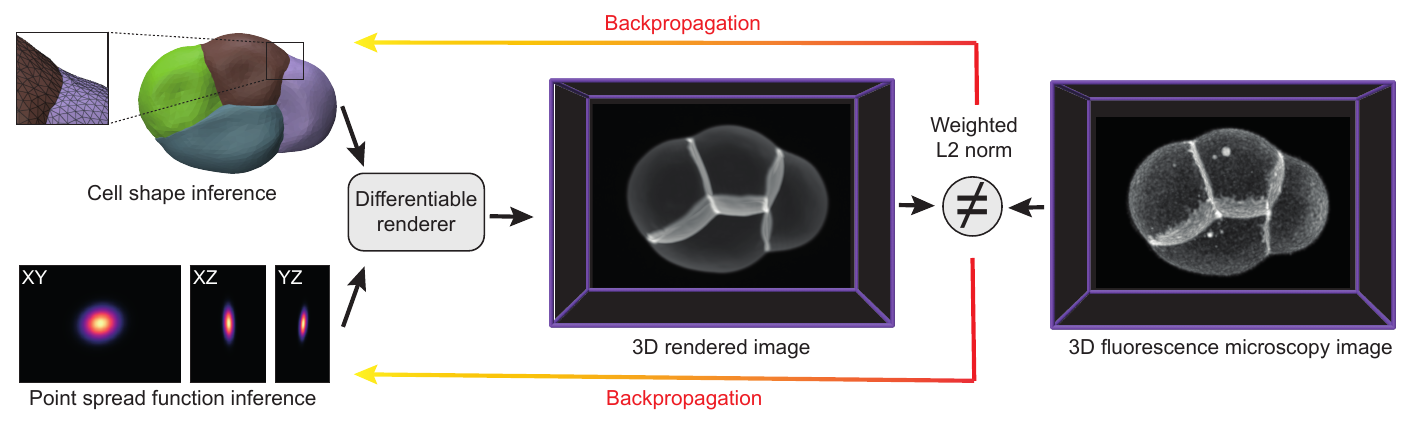}
    \caption{Graphical summary: joint inference of cell geometry and optical PSF using our differentiable 3D fluorescence microscopy.}
\end{figure*}
Fluorescence microscopy~\cite{lichtman2005fluorescence} is the most widely used technique for imaging biological structures. However, extracting quantitative information from 3D fluorescence images remains a major bottleneck, limiting the development of new biological shape analysis methods and models. In fluorescence microscopy, biological samples are labeled with fluorescent dyes (fluorophores) that bind to specific structures of interest—such as lipid membranes, cytoskeletal networks, or organelles—making them bright while keeping the background dark. Except for unavoidable camera noise and biological heterogeneities, 3D fluorescent images are typically sparse and composed of well-structured objects such as points, filaments, surfaces, or bulk solids.

Leveraging this inherent structure, tools based on deep-learning (DL) tailored for fluorescence imaging~\cite{belthangady2019applications}, particularly convolutional neural networks (CNNs)~\cite{10.1007/978-3-319-24574-4_28,10.1007/978-3-319-46723-8_49}, have enabled significant advances in automating challenging tasks such as image restoration~\cite{CARE}, instance segmentation~\cite{schmidt2018,Cellpose}, and feature encoding via self-supervised learning~\cite{cytoself}.
Despite these achievements, the expressivity of neural networks also presents challenges in 3D biological imaging:
(1) Data requirements: 3D CNNs or newer architectures (e.g., Vision Transformers~\cite{dosovitskiy2020image}) require large annotated datasets, which are difficult to produce and label accurately in 3D. While foundation models for biomedical image analysis show promise~\cite{ma2024segment}, commonly used DL models remain highly dependent on specific microscope imaging modalities, requiring extensive retraining when imaging conditions change.
(2) Lack of interpretability and physical constraints: widely-used DL segmentation models~\cite{Cellpose} lack interpretability and rarely incorporate prior knowledge about optical image formation, limiting their scientific insight.
(3) Geometric analysis: CNNs typically produce segmentation masks, which are suboptimal for measuring geometric features like curvatures or angles~\cite{ichbiah2023embryo}. Additionally, timelapse microscopy requires object shape tracking~\cite{uhlmann2017flylimbtracker}, yet most segmentation models operate frame by frame, necessitating a separate shape-matching step as post-processing.
(4) Risk of hallucinations: DL models can generate artifacts or structures that do not correspond to real biological features—an issue for scientific research and medical applications, where accuracy is critical.

In response to these challenges, a growing trend in microscopy image analysis favors methods that incorporate stronger prior knowledge of optical physics~\cite{DBLP:conf/icml/BatsonR19,DBLP:conf/icml/LehtinenMHLKAA18,ahmet_can_solak_2022_7222198,banerjee2024physics}. In contrast to DL, active contour methods—also known as active snakes or meshes—directly evolve a shape representation to minimize an objective function aligned with image features~\cite{chanvese,DBLP:journals/ijcv/KassWT88,dufour2005segmenting,limeseg,dufour20103,uhlmann2016hermite,smith2023active}. Although being developed for decades, these methods have seldom incorporated the physical principles of fluorescence image formation, instead relying on user-defined regularization terms to enforce smoothness. While DL approaches recently supplanted them due to superior speed and automation, active contour methods retain valuable advantages: they can track objects over time and produce explicit shape representations (e.g., meshes, level sets) that ease geometric analysis and integrate naturally with downstream modeling frameworks.

Meanwhile, the emergence of libraries such as PyTorch~\cite{torch} and JAX~\cite{jax}, which offer native automatic differentiation, has accelerated the adoption of differentiable models in computer vision (CV), notably in inverse rendering~\cite{kato2020differentiable,Nicolet2021Rendering} and mesh-based shape inference~\cite{NEURIPS2024_1651f1ea}. Although inverse rendering has proven effective in CV, it remains largely unexplored in fluorescence biological image analysis. Here, we introduce a novel paradigm that leverages the fundamental optics of fluorescence microscopy to approximate the 3D image formation process from biological object shapes. We present \textit{deltaMic}, a differentiable model that renders 3D fluorescence microscopy images from surface meshes, which can be directly compared to real biological images using a weighted voxel-based $L^2$ norm. Implemented in PyTorch for automatic differentiation, our approach efficiently optimizes this loss by computing gradients with respect to both the biological object’s shape and the microscope’s optical properties. As a result, our method enables precise biological shape inference while simultaneously emulating the microscope’s optics, eliminating the need for per-sample hyperparameter tuning required in traditional active contour methods.

\noindent \textbf{Main contributions:} \\
$\bullet$ We introduce a simplified model of 3D fluorescence microscopy generation, combining a mesh-based object representation with a parameterized point spread function (PSF). \\
$\bullet$ We implement a differentiable Fourier transform of triangle surface meshes parallelized on GPU.\\
$\bullet$ We demonstrate that our differentiable microscopy renderer can accurately infer 3D cellular shapes from both synthetic and real 3D microscopy images without requiring additional shape-regularization terms. \\
$\bullet$ We provide a PyTorch implementation on \href{https://github.com/VirtualEmbryo/deltaMic}{GitHub} \cite{ichbiah_2025_16623962}.

\section{Related work}

\subsection{Fluorescence microscopy modeling}
In fluorescence microscopy, a laser excites fluorescent dyes bound to biological structures at a specific wavelength, causing them to emit light at a longer wavelength. The emitted fluorescence passes through the optics and is captured by a CMOS camera, producing a 2D image. In modalities such as confocal or light-sheet microscopy, the focal plane is incrementally shifted to acquire a 3D volumetric image composed of optical sections~\cite{keller2013imaging}. However, resolution is fundamentally limited by the diffraction of light through the optical system~\cite{Inoue2006}. When the imaging system’s response is approximately translation-invariant in 3D, it can be described by a point spread function (PSF), which characterizes how a point source is imaged. Various physics-based PSF models have been proposed~\cite{Gibson:92,Kraus:89,Hanser:03,Hanser:04,ARNISON200253}, mainly for deconvolution purposes~\cite{sage2017deconvolutionlab2}. The simplest way to determine a PSF is to image sub-micrometer fluorescent beads approximating point sources. The model is then fitted to these experimental images. In this work, we approximate the PSF using a basic Gaussian kernel for simplicity; however, the framework could also incorporate more accurate experimental or physics-based models~\cite{aguet2009super}.

Knowing the PSF also enables synthetic image generation, explored in several studies. These approaches either replicate the real image-formation process~\cite{dmitrieff2017confocalgn} or use texture synthesis~\cite{malm2015simulation,wiesmann2017using} or generative DL~\cite{HOLLANDI2020453,eschweiler20213d} to produce realistic images. Such images are primarily used to evaluate image-analysis algorithms~\cite{rajaram2012simucell} or generate annotated datasets for DL training~\cite{mill2021synthetic}. In~\cite{dmitrieff2017confocalgn}, the fluorophore distribution is modeled as a Boolean mask convolved with a user-provided PSF and noise. We extend this approach by defining a fluorophore density over each mesh simplex rather than using discrete sources or a mask. This avoids densely populating biological meshes with millions of points and significantly reduces computation.

\subsection{Instance segmentation of fluorescent images}
Biological image segmentation has evolved in parallel with computer vision~\cite{6279591}, from traditional methods such as thresholding, watershed~\cite{beucher1992watershed,beucher2018morphological,fernandez2010imaging}, and active contours~\cite{DBLP:journals/ijcv/KassWT88, chanvese, limeseg, dufour20103}, to graph-cut optimization \cite{boykov2006graph}, and more recently DL-based pipelines~\cite{10.1007/978-3-319-24574-4_28,10.1007/978-3-319-46723-8_49}. In 2D, the most effective methods leverage large annotated datasets to train CNNs~\cite{Cellpose2,cutler2022omnipose,TissueNet} or, more recently, vision Transformers~\cite{ma2024segment}. These models predict instance masks robustly across samples and imaging modalities. However, extending this success to 3D remains challenging.

3D images (z-stacks) are acquired by capturing 2D slices at varying depths, introducing anisotropy along the z-axis that varies with imaging conditions, yielding heterogeneous datasets that hinder CNN training and generalization. Modern 3D images—ranging from $512^3$ voxels in confocal to $2048^3$ in light-sheet microscopy—also impose substantial memory constraints on GPUs. Labeling 3D data poses major challenges. It requires expert biological knowledge, strong spatial skills to maintain cross-plane consistency, and suitable annotation tools. The shift from 2D to 3D increases both the number of voxels and annotation complexity.

Classical contour-based segmentation techniques often minimize an energy functional $\mathcal{E}(\Lambda,m)$ modeling the distance between a shape $\Lambda$ and image features $m$~\cite{mumford1989optimal}, such as intensity gradients~\cite{malladi1995shape} or homogeneous regions~\cite{chanvese}. Shape representations use level sets or meshes, and the optimal shape $\Lambda^*$ is found via gradient-based optimization. The negative gradient $\frac{\partial \mathcal{E}(\Lambda,m)}{\partial \Lambda}$ acts as a force guiding the contour—hence the terms active snakes, contours~\cite{DBLP:journals/ijcv/KassWT88,limeseg,uhlmann2016hermite}, or meshes~\cite{dufour20103,smith2023active}. These methods typically require user-defined regularization (e.g., penalization terms akin to tension or bending energies) to maintain smoothness and avoid artifacts.

Fluorescence images of slender structures like membranes are challenging for region- or edge-based energies, as such structures appear as thin, high-intensity features against a dark background. Alternatives use forces based on proximity to local intensity maxima~\cite{limeseg,10.1093/bioinformatics/btab557} or compare real and synthetic images generated from meshes~\cite{10.1007/978-3-642-10331-5_51,Nguyen2016}, where the synthetic image is produced by convolving a binary mesh mask with a known PSF. Our approach builds on this idea but introduces a more rigorous 3D rendering framework and direct variational formulation. We jointly optimize the mesh and PSF parameters in a differentiable pipeline, eliminating the need for explicit shape regularization while maintaining robust and physically grounded reconstructions.

\subsection{Differentiable rendering}
Rendering 2D or 3D geometrical shapes into rasterized images is a core problem in computer graphics. Recent advances have enabled differentiable rendering, allowing inverse problems where scene parameters (e.g., shapes, textures, materials) are learned from raster images~\cite{10.1145/3414685.3417871,10.1145/3355089.3356498,DBLP:journals/corr/abs-2006-12057}. Traditional rasterization is non-differentiable due to occlusions and discontinuities, prompting development of smoothing strategies~\cite{pytorch3d,Liu_2019_ICCV,10.1145/3414685.3417861,DBLP:conf/cvpr/KatoUH18}.

Our fluorescence microscopy emulator functions as a 3D differentiable renderer. Like in graphics, it relies on mathematically differentiable operations. Unlike graphics, the resulting image lives in 3D space. Image formation is smoothed by convolving the mesh with a PSF, similarly to smoothing in differentiable rasterization~\cite{Liu_2019_ICCV}. Since a triangle mesh may include thousands of vertices and a PSF hundreds of parameters, optimization demands efficient differentiation. Assuming all components are differentiable, reverse-mode differentiation (backpropagation) computes gradients via the chain rule. Growing demands for scalability have led to GPU-accelerated automatic differentiation libraries~\cite{torch,jax}. We leverage PyTorch’s backpropagation engine and implement our mesh Fourier Transform (FT) as a differentiable function within it.

\begin{figure}[h]
  \centering
  \includegraphics[width=\linewidth]{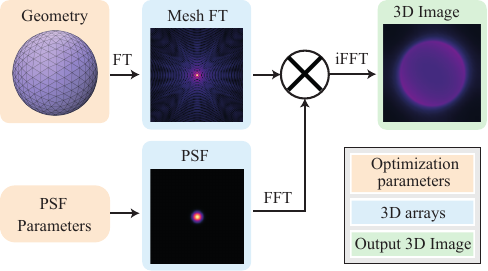}
  \caption{\textit{DeltaMic} pipeline overview: The microscopy renderer generates a 3D synthetic fluorescence image from a triangle mesh and a parameterized PSF emulating microscope optics. Gradients with respect to mesh and PSF parameters are computed via backpropagation, enabling their optimization to match real microscopy images.}
  \label{fig:pipeline_overview}
\end{figure}

\section{Fluorescent 3D image rendering}
Images can be represented as intensity maps $I$ from $[0,1]^3$ to $\mathbb{R}$ without loss of generality, as any non-cubic image can be linearly rescaled to fit within $[0,1]^3$. Following subsections introduce approximated models for both: imaging system and the geometry of biological samples, that are then integrated to construct our differentiable renderer.
\begin{figure}[h]
  \centering
  \includegraphics[width=\linewidth]{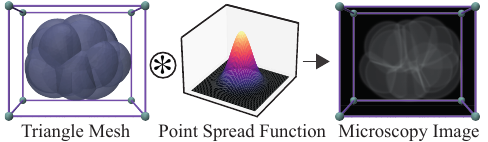}
  \caption{Principle of 3D fluorescence microscopy rendering: A fluorophore distribution on the cell surface $\Lambda$ is convolved with the system’s PSF to produce a 3D fluorescence image.}
\end{figure}

\subsection{Translation-invariant rendering model}
We assume that the PSF  $h$  of the fluorescence microscope is translation-invariant. Under this assumption, a smooth image  $I: [0,1]^3 \to \mathbb{R}$  is obtained by convolving a spatial fluorophore density  $u_\Lambda: [0,1]^3 \to \mathbb{R}$  with the point spread function (PSF) kernel  $h: [0,1]^3 \to \mathbb{R}$ :
\begin{equation}
I(\textbf{x}) = (u_{\Lambda} *  h)(\textbf{x}) = \int_{[0,1]^3}  u_\Lambda (\textbf{p}) h(\textbf{x}-\textbf{p}) \mathrm{d}^3\textbf{p}.
\end{equation}
To circumvent the direct computation of these integrals, we perform the convolution in Fourier space by applying an element-wise multiplication of the Fourier transform (FT) $\hat{u}_{\Lambda}$ and $\hat{h}$, yielding: $\hat{I} = \hat{u}_{\Lambda} \cdot \hat{h}.$

With this image formation model, one can start from a first guess $\left(I^0,h^0\right)$, and minimize the distance between the rendered image $I^0$ and a real microscopy image $I_{\text{GT}}$. Doing so allows to infer both the fluorophore density of the observed biological sample and the PSF of the imaging system: $\left(I^*,h^*\right)$. However, to learn meaningful shape representations, both of these functions have to be parameterized using prior knowledge on the shape considered, its topology, as well as on the PSF.

\subsection{Point spread function model}
The simplest PSF model is a Gaussian kernel, which is is fully determined by its covariance matrix $\Sigma \in \mathbb{R}^{3\times3}$: 
\begin{equation}
    h(\mathbf{x})= \dfrac{\mathrm{e}^{-\frac{1}{2}\mathbf{x}^T\Sigma^{-1}\mathbf{x}}}{\sqrt{(2\pi)^3\det\Sigma}}\;,\;\mathbf{x} \in\mathbb{R}^3.
\end{equation}
The FT $\hat{h}$ of $h$ is given by: $\hat{h}(\boldsymbol{\xi}) = \mathrm{e}^{-\frac{1}{2}\boldsymbol{\xi}^T\Sigma^{-1}\boldsymbol{\xi}},$ where $\boldsymbol{\xi} \in \mathbb{R}^3$ is the wavevector.
In this framework, inferring the PSF consists of optimizing the coefficients of the symmetric positive semi-definite covariance matrix $\Sigma$, which directly controls the level of blur in the rendered image. More photorealistic PSFs based on differentiable physical models could also be incorporated~\cite{Hanser:03,Hanser:04}, allowing the learned parameters to correspond to real optical properties, such as refractive indices or optical aberrations expressed using Zernike polynomials~\cite{Zernikereview}.

\subsection{Geometric models of biological objects}
Extracting geometry from volumetric images involves approximating biological structures using discrete $N$-dimensional representations ($N = 0,1,2,3$) embedded in 3D space. These structures vary in size and topology. We focus here on cell membranes, modeled as 2D surfaces using triangle meshes, though extensions to points, filaments, and volumes are possible. This choice reflects prior knowledge of structural topology: as shown in Section~\ref{celegans_ex}, early embryos comprise cells forming bounded regions, representable as a single non-manifold, multimaterial mesh~\cite{Multitracker,maitre2016asymmetric}, where cell–cell interfaces are captured by doubling intensity at overlapping membranes.

\section{Fourier transform of triangulated surfaces}
We derive an explicit expression for the FT of arbitrary 2D surfaces embedded in $\mathbb{R}^3$, and provide closed-form formulas for both the FT and its gradient with respect to vertex positions in the case of a triangulated mesh.

\subsection{Surfaces as spatial Dirac distributions}
We model the spatial fluorophore density $u_\Lambda$ as a Dirac distribution supported on the surface $\Lambda$, weighted by the local surface element to reflect the actual fluorophore density along the surface.

\begin{figure}[h]
  \centering
  \includegraphics[width = 0.7\linewidth]{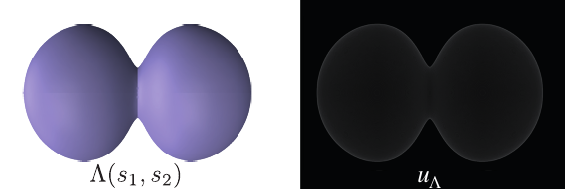}
  \caption{Surface $\Lambda$ describing a dividing cell and its associated fluorophore density $u_\Lambda$ convolved with a Gaussian PSF in $\mathbb{R}^3$ (maximum projection).}
\end{figure}

We consider a surface parametrized by a function $\boldsymbol{\Lambda}:\,[0,1]^2 \to [0,1]^3,$ mapping local coordinates $(s_1,s_2)$, defined in a tangent orthonormal basis, to points in 3D space. The spatial fluorophore density in $\mathbb{R}^3$ is then defined as: 
\begin{equation}
u_{\Lambda}(\mathbf{x}) = \dfrac{1}{|\Lambda|}\iint \delta_{\boldsymbol{\Lambda}(s_1,s_2)}(\mathbf{x})\,a_{\Lambda}(s_1,s_2)\,\mathrm{d}s_1\mathrm{d}s_2,
\end{equation}
where $ \delta_{\boldsymbol{\Lambda}(s_1,s_2)}$ denotes the Dirac delta function centered at $\boldsymbol{\Lambda}(s_1,s_2)$, $a_{\Lambda}\!=\!\|\partial_{s_1}\boldsymbol{\Lambda}\!\times\!\partial_{s_2}\boldsymbol{\Lambda}\|$ is the surface element. The normalization by the total surface area $|\Lambda|=\iint a_{\Lambda}(s_1,s_2)\mathrm{d}s_1\mathrm{d}s_2$ ensures that the total integrated image intensity is one. When dealing with several surfaces with different densities, one can use a weighted sum or integral of $u_{\Lambda}$. Cell-cell interfaces are modeled as double membranes, and their fluorophore density is therefore scaled by a factor two.

For all $\mathbf{x}\!=\!(x,y,z) \in \mathbb{R}^3$, the FT of the Dirac distribution at a point $\mathbf{x}$ is given by $\hat{\delta}_{\mathbf{x}}(\boldsymbol{\xi}) = \mathrm{e}^{-\mathrm{i} \mathbf{x} \cdot \boldsymbol{\xi}}, \, \boldsymbol{\xi} \in \mathbb{R}^3.$ By linearity, the FT of $u_{\Lambda}$ reads therefore:
\begin{equation}
\hat{u}_{\Lambda}(\boldsymbol{\xi}) = \dfrac{1}{|\Lambda|}  \iint  a_{{\Lambda}}(s_1,s_2) \,\mathrm{e}^{-\mathrm{i} \boldsymbol{\Lambda}(s_1,s_2)\cdot\boldsymbol{\xi}} \,\mathrm{d}s_1 \mathrm{d}s_2.
\end{equation}

In practice, we discretize $\Lambda$ with triangle meshes. By linearity, this reduces to computing and summing the FT of each triangle.

\subsubsection{Spatial density for a triangle mesh}
A triangle mesh is a surface $\Lambda = \{\mathcal{T}\}$ defined by a set of triangles. For $\boldsymbol{\xi} \in \mathbb{R}^3$, by linearity, its FT is defined by: 

\begin{equation}
    \hat{u}_\Lambda(\boldsymbol{\xi}) = \frac{\underset{\mathcal{T} \in \Lambda}{\sum} \hat{u}_\mathcal{T} (\boldsymbol{\xi})}{|\Lambda|} = \frac{\underset{\mathcal{T} \in \Lambda}{\sum} \hat{u}_\mathcal{T} (\boldsymbol{\xi})}{\underset{\mathcal{T} \in \Lambda}{\sum} A_\mathcal{T}},
\end{equation}
Where $A_\mathcal{T}$ is the area of $\mathcal{T}$. The gradient of the FT with respect to a vertex $\textbf{v}$ reads: 
\begin{equation}
    \frac{\partial \hat{u}_\Lambda(\boldsymbol{\xi})}{\partial \textbf{v}}  = \frac{1}{|\Lambda|}\left( 
    \underset{\mathcal{T} \in \textbf{v}\star}{\sum} \frac{\partial \hat{u}_\mathcal{T} (\boldsymbol{\xi})}{\partial \textbf{v}} - 
    \hat{u}_\Lambda (\boldsymbol{\xi}) \underset{\mathcal{T} \in \textbf{v}\star}{\sum} \frac{\partial A_\mathcal{T}}{\partial \textbf{v}}
    \right),
\end{equation}
where $\textbf{v}\star$ denotes the set of the triangles of $\Lambda$ that contains the vertex $\textbf{v}$.

\begin{figure}[h]
  \centering
  \includegraphics[width=0.7\linewidth]{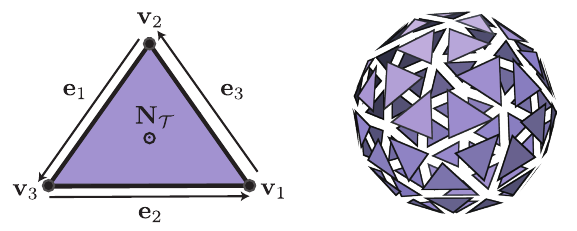}
  \caption{Notations for a triangle mesh.}
\end{figure}

\subsubsection{Fourier transform and gradients of a triangle}
We consider a triangle $\mathcal{T}$ defined by its vertices $(\textbf{v}_1,\textbf{v}_2,\textbf{v}_3)$ and compute $\hat{u}_\mathcal{T}$ and its spatial derivatives. For convenience, we set $\textbf{v}_4 \!=\! \textbf{v}_1$ and similarly by permutation $\textbf{v}_0 \!=\! \textbf{v}_3$. For $p\!=\!1 \ldots 3$, we define $p^- \!=\!p\!-\!1$ and $p^+\!=\! p\!+\!1$ and denote by $\textbf{e}_p = \textbf{v}_{p^-} \!-\! \textbf{v}_{p^+}$  the opposite edge to $\textbf{v}_p$ and by $l_p = |\textbf{e}_p|$ its length.
$A_\mathcal{T} \!=\! \dfrac{|\textbf{e}_3 \times \textbf{e}_1|}{2}$ denotes the area of the triangle, and $\textbf{N}_\mathcal{T} \!=\! \dfrac{\textbf{e}_3 \times \textbf{e}_1 }{2A_\mathcal{T}}$ denotes its unit normal. For $p\!=1\! \ldots 3$, we define $\textbf{w}_p = \textbf{e}_p \times \textbf{N}_{\mathcal{T}}$, the non-normalized outward normal to $\mathcal{T}$ on the edge $\textbf{e}_p$. This allows us expressing the gradient of the triangle area with respect to each vertex position as $\dfrac{\partial A_\mathcal{T}}{\partial \textbf{v}_p}  = - \dfrac{\textbf{w}_p}{2}$, for $p=1\ldots3$.

Expressing the FT of the density on a triangle as $\hat{u}_{\mathcal{T}}(\boldsymbol{\xi}) = \int_\mathcal{T} \mathrm{e}^{-\mathrm{i}\mathbf{x} \cdot \boldsymbol{\xi}} \mathrm{d}s(\mathbf{x})$, then for all $\boldsymbol{\xi} \in \mathbb{R}^3$, we have: 
\begin{align}\label{fsum}
\hat{u}_{\mathcal{T}}(\boldsymbol{\xi}) &= 2A_\mathcal{T} f_\mathcal{T}(\boldsymbol{\xi}), \\
    \text{with} \quad f_\mathcal{T}(\boldsymbol{\xi}) &= \sum_{p=1}^3 \frac{\mathrm{e}^{-\mathrm{i}\mathbf{v}_p\cdot \boldsymbol{\xi}}}{(\textbf{e}_{p^-}\cdot \boldsymbol{\xi})(\textbf{e}_{p^+}\cdot \boldsymbol{\xi})}.
\end{align}
We deduce 
\begin{equation}
    \frac{\partial \hat{u}_{\mathcal{T}}}{\partial \textbf{v}_p} (\boldsymbol{\xi}) = -f_\mathcal{T}(\boldsymbol{\xi})\textbf{w}_p +  2A_\mathcal{T} \frac{\partial f_\mathcal{T}(\boldsymbol{\xi})}{\partial \textbf{v}_p},\, p=1\ldots3
\end{equation}
\begin{equation}
    \begin{split}
    \text{with}\,\dfrac{\partial f_\mathcal{T}(\boldsymbol{\xi})}{\partial \textbf{v}_p}=  & \boldsymbol{\xi}\left[\dfrac{\mathrm{e}^{-\mathrm{i} \mathbf{v}_{p^+}\cdot\boldsymbol{\xi}}}{(\textbf{e}_{p^-}\cdot\boldsymbol{\xi})^2(\textbf{e}_p\cdot\boldsymbol{\xi})}-\dfrac{\mathrm{e}^{-\mathrm{i} \mathbf{v}_{p^-}\cdot\boldsymbol{\xi}}}{(\textbf{e}_p\cdot\boldsymbol{\xi})(\textbf{e}_{p^+}\cdot\boldsymbol{\xi})^2}\right. \\
 & \left.-\dfrac{\mathrm{i}\mathrm{e}^{-\mathrm{i} \mathbf{v}_p\cdot\boldsymbol{\xi}}}{(\textbf{e}_{p^-}\cdot\boldsymbol{\xi})(\textbf{e}_{p^+}\cdot\boldsymbol{\xi})}
+\dfrac{\mathrm{e}^{-\mathrm{i} \mathbf{v}_p\cdot\boldsymbol{\xi}}}{(\textbf{e}_{p^-}\cdot\boldsymbol{\xi})^2(\textbf{e}_{p^+}\cdot\boldsymbol{\xi})}\right.\\
&\left.-\dfrac{\mathrm{e}^{-\mathrm{i} \mathbf{v}_p\cdot\boldsymbol{\xi}}}{(\textbf{e}_{p^-}\cdot\boldsymbol{\xi})(\textbf{e}_{p^+}\cdot\boldsymbol{\xi})^2}\right].
    \end{split}
\end{equation}

\subsubsection{Numerical approximations to prevent divergence}
The Fourier Transform (FT) of a Dirac distribution on a triangle is $C^\infty$. However, significant computational errors arise when a denominator in Eq.~\eqref{fsum} approaches zero. Due to rounding errors in floating-point arithmetic, numerical precision is limited to a threshold $\epsilon$, below which values cannot be reliably distinguished from zero.
To mitigate this issue, when a denominator term in Eq.~\eqref{fsum} approaches zero, we substitute it with a stable approximation, which we detail in the following section.  We write $f_\mathcal{T}(\boldsymbol{\xi})=g(\textbf{e}_1\cdot\boldsymbol{\xi},\textbf{e}_2\cdot\boldsymbol{\xi},\textbf{e}_3\cdot\boldsymbol{\xi})$, with the function $g$ defined for $(s,t,u)\in\mathbb{R}^3$ by
\begin{equation}\label{gstu}
\begin{array}{ll}
     g(s,t,u)\!= \!\dfrac{-\,\mathrm{e}^{i s}}{(s\!-\!t)(s\!-\!u)}
     \!+\!\dfrac{-\,\mathrm{e}^{i t}}{(t\!-\!u)(t\!-\!s)}
     \!+\!\dfrac{-\,\mathrm{e}^{i u}}{(u\!-\!s)(u\!-\!t)}.
\end{array}
\end{equation}
When two values $(a, b)$ among $(s, t, u)$ satisfy $|a \!-\! b| < \epsilon$, the denominator in Eq.~\eqref{gstu} approaches zero, leading to divergence. To prevent this, we derive exact expressions for $g(s,t,u)$ in the special cases where  $t = u ,  t = s ,  u = s $, or  $u = s = t$ . These alternative expressions replace the original formulation in Eq.~\eqref{gstu} whenever $|a \!-\! b| < \epsilon$, ensuring numerical stability:
\begin{subequations}
\begin{align}
     g(t,t,u)&=g(u,t,t) = g(t,u,t) \nonumber \\
     &=i\dfrac{\mathrm{e}^{-\mathrm{i} t}}{t-u}+\dfrac{\mathrm{e}^{-\mathrm{i} t}}{(t-u)^2}-\dfrac{\mathrm{e}^{-\mathrm{i} u}}{(t-u)^2}.\\
     g(u,u,u)&=\frac{\mathrm{e}^{-\mathrm{i} u}}{2}.
\end{align}    
\end{subequations}

\section{Numerical implementation}
\subsection{Acceleration}
Computing the artificial image requires evaluating the FT $\hat{u}_\Lambda$ for each of the $N$ voxels times the number $n_t$ of triangles in the mesh, resulting in a runtime complexity of $\mathcal{O}(N \cdot n_t)$. The gradient of the mesh FT involves an $\mathcal{O}(1)$ sum per voxel for each of the $n_v$ vertices, leading to a total complexity of $\mathcal{O}(N \cdot n_v)$.
In practice, a confocal microscopy image of size $500^3$ contains 125 million voxels, while a reasonable mesh typically consists of $10^{3-4}$ vertices and triangles. Without proper parallelization, these computations become prohibitive. The high computational cost has been a key limitation preventing the widespread use of spectral methods in 3D, despite their promising applications~\cite{jiang2019convolutional}.

We propose two complementary strategies to accelerate the mesh FT: GPU parallelization and a narrow-band approximation method in the frequency domain.

\begin{figure}[h!]
  \centering
  \includegraphics[width=\linewidth]{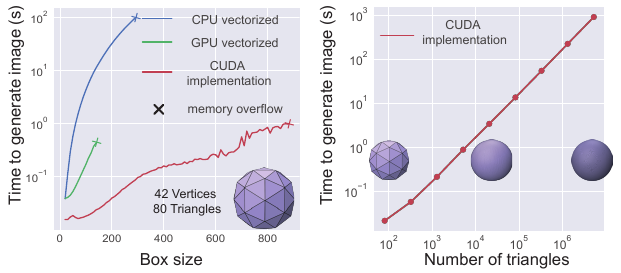}
  \caption{Benchmarking of Mesh FT Computation: (Left) For a 80-triangle spherical mesh, our CUDA implementation significantly improves speed and memory efficiency as function of image size. (Right) For an image of size $[100]^3$, the forward pass scales linearly with the number of triangles.}
  \label{fig:figbench}
\end{figure}

\noindent\textbf{GPU parallelization}
The grid-based structure of a 3D image naturally lends itself to massively parallel computations on graphical processing units (GPUs). To leverage this, we provide a custom CUDA implementation for both, the forward and the backward passes. As shown in Figure~\ref{fig:figbench}, our implementation achieves  $\approx10^3$ speedup on one NVIDIA V100 GPU compared to a vectorized CPU implementation, while also enabling computations on significantly larger box sizes before encountering overflow.

\noindent\textbf{Narrow-band approximation in the frequency domain}
As previously stated, the FT of the image is given by the element-wise product of the FT of the mesh and the PSF:  $\hat{I} = \hat{u}_{\Lambda} \cdot \hat{h}$. In practice, the PSF acts as a low-pass filter, emphasizing low spatial frequencies of the mesh. A blurred image corresponds to a sparse PSF, with its highest amplitudes concentrated near zero frequency, thereby suppressing fine mesh details. The more resolved the original biological image, the more frequency components of the PSF are required for accurate rendering. For any given spatial frequency $\boldsymbol{\xi}$, if $\hat{h}(\boldsymbol{\xi}) \approx 0$, then computing $\hat{u}_{\Lambda}(\boldsymbol{\xi})$ is unnecessary, as it contributes negligibly to $\hat{I}(\boldsymbol{\xi})$ and hence $I(\mathbf{x})$. To exploit this, we apply a frequency-domain cutoff, computing the FT of the mesh only for high-frequency, low-amplitude regions of the PSF (only where its FT exceeds $1\%$ of its maximum value). This spectral narrow-band method reduces computational cost by several orders of magnitude, skipping computations with negligible far-field contributions in real space. 

\noindent\textbf{A staggered optimization scheme}\label{strategies}
Despite significant speed improvements, computing the mesh FT and its gradient remains the main computational bottleneck of our pipeline. We offer two approaches for computing the FT: a fast, approximate method using the narrow-band approach and a slower, exact method performing the full computation. The forward pass is approximately 10 times slower than the backward pass, making it crucial to apply the narrow-band method in the backward computation whenever possible. To address this, we decompose the coupled optimization problem into two staggered sub-problems, iterating them sequentially.
\noindent (a) Shape optimization: To optimize the shape, an approximate computation of the mesh FT suffices. We use the narrow-band method to efficiently compute the FT and perform an optimization step for the vertex positions while keeping the PSF fixed.
\noindent (b) PSF optimization: Optimizing the PSF requires computing the mesh FT across all available frequencies, avoiding bias from the narrow-band threshold, which depends on the current PSF values. Thus, we compute the exact FT of the mesh and perform an optimization step for the PSF parameters while keeping the shape fixed.

We optimize shapes by minimizing a weighted $L^2$ loss $\mathcal{L} = \left< (I - I_{\text{GT}})^2, I_{\text{GT}} \right>$, where $\left<\cdot,\cdot\right>$ denotes the Frobenius inner product. This formulation downweights dark regions in the ground-truth image $I_{\text{GT}}$ to focus the optimization on informative signal. Following~\cite{Nicolet2021Rendering}, we update mesh vertex positions using the AdamUniform optimizer and apply a diffusion-based regularization to the gradient:
$\mathbf{v} \leftarrow \mathbf{v} - \eta\, (\boldsymbol{\mathds{1}} + \lambda \mathbf{L})^{-2} \frac{\partial \mathcal{L}}{\partial \mathbf{v}}$, where $\mathbf{L} \in \mathbb{R}^{n_v \times n_v}$ is the mesh Laplacian and $\lambda = 50$. Computing $(\boldsymbol{\mathds{1}} + \lambda \mathbf{L})^{-2}$ explicitly is memory-intensive for large vertex counts $n_v$, so we instead solve the equivalent linear system $(\boldsymbol{\mathds{1}} + \lambda \mathbf{L})^{-2} \mathbf{A} = \frac{\partial \mathcal{L}}{\partial \mathbf{v}}$ via GPU-parallelized sparse Cholesky decomposition~\cite{naumov2011parallel,Nicolet2021Rendering}.

\subsection{PSF optimization step}
PSF parameter optimization aims to accurately reconstruct both bright and dark regions of the image. A regular $L^2$ norm minimization,
$\mathcal{L}_0 = ||I-I_{\text{GT}}||^2 $ is well-suited for this task. In this step, we optimize the covariance matrix $\Sigma$ using the Adam optimizer~\cite{kingma2014adam}.

\section{Benchmarking and application}

\subsection{Benchmarks on artificial images}
\begin{figure}[h!]
  \centering
  \includegraphics[width=\linewidth]{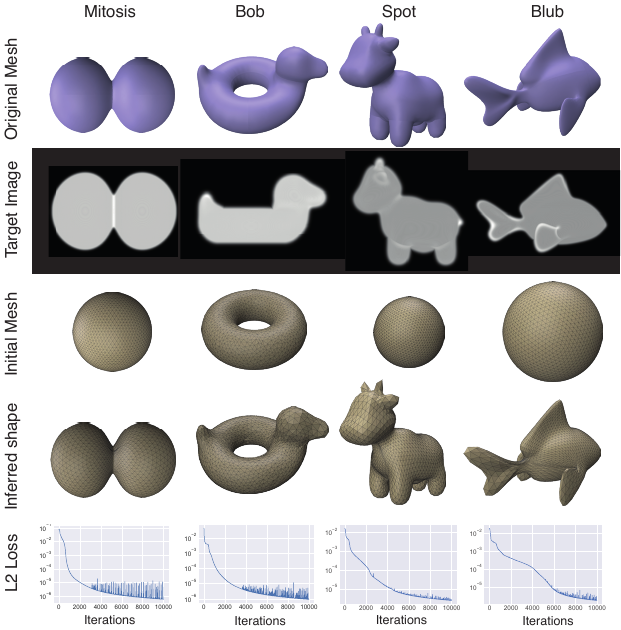}
  \caption{Shape inference from 3D artificial images: Starting from spherical or toroidal meshes, the optimization process reconstructs meshes that closely align with the original shapes (see also: Supplemental videos 1–4).}
  \label{fig:artificial_examples}
\end{figure}

We first benchmark our method on artificial images generated from common computer graphics meshes and a simulated dividing cell (Figure~\ref{fig:artificial_examples}). Images are rendered using a Gaussian isotropic PSF, which was also used for inference to isolate the effect of shape reconstruction. Starting from elementary mesh shapes (a sphere or torus, depending on topology), we perform 10,000 optimization iterations per example. Despite the absence of remeshing or collision detection algorithms~\cite{brochu2012efficient,lin2017collision}, our regularized shape optimization accurately converges to the original shapes.

\begin{figure}[h!]
  \centering
  \includegraphics[width=0.98\linewidth]{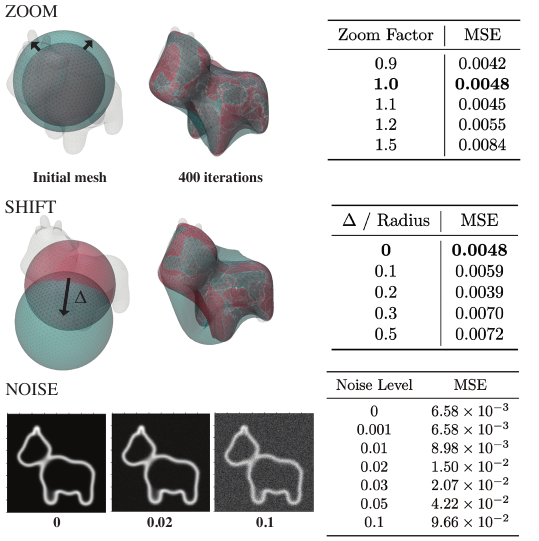}
  \caption{Benchmarking robustness to initialization and noise. (Top, Middle) MSE evaluation for shape reconstruction using Spot as the reference mesh, with variations in initial mesh scale and position. Larger shifts increase convergence time. (Bottom) The method is quite robust to random noise.}
  \label{fig:benchmark}
\end{figure}

\subsection{Robustness to initial conditions and noise}
Next, we evaluate the convergence robustness to initial conditions and noise. As a metric, we compute the mean-squared error (MSE) between the reconstructed shape—rendered as an artificial image—and the target image generated from the initial mesh. Using Spot (cow) as the reference shape, we assess the MSE when the starting mesh is scaled (zoomed in/out) and shifted from its barycenter (Figure~\ref{fig:benchmark}, top and middle). While our method is generally robust to initialization, large shifts from the target increase the number of iterations needed for convergence. Notably, our approach demonstrates high robustness to random noise added to the image (Figure~\ref{fig:benchmark}, bottom), which is particularly advantageous for microscopy applications.

\begin{figure}[h]
 \centering
  \includegraphics[width=\linewidth]{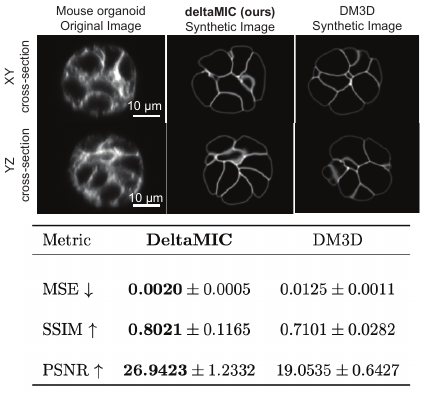}
  \caption{Benchmarking \textit{deltaMic} vs DM3D~\cite{smith2023active} on 3D mouse organoid images.}
  \label{fig:comparisonDM3D}
\end{figure}

\begin{figure*}[h!]
  \centering
  \includegraphics[width=\textwidth]{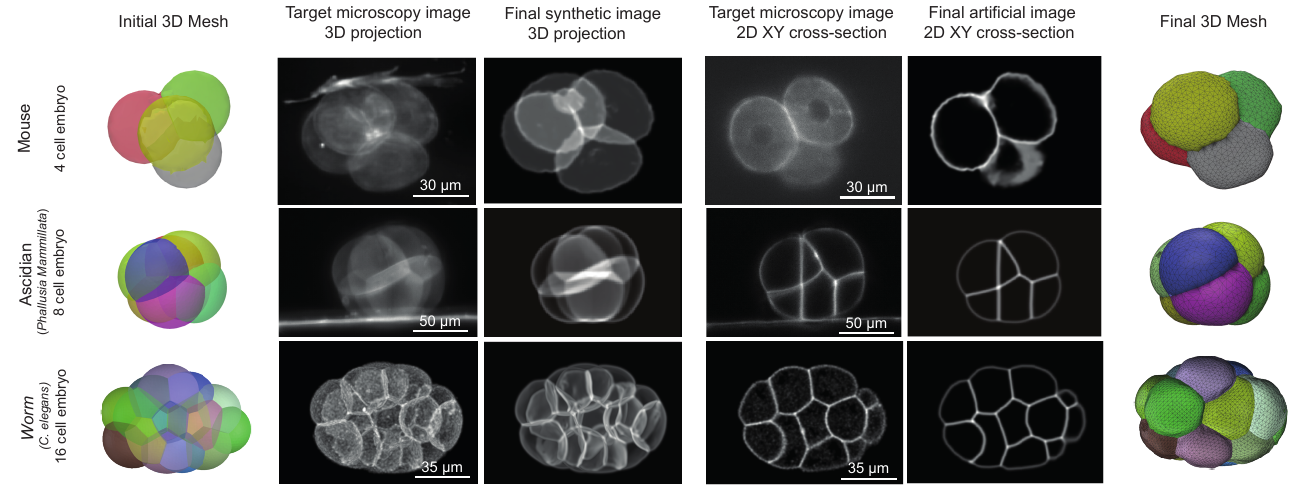}
  \caption{Cell shape inference in early embryos: Starting from a foam-like mesh, both vertex positions and PSF are optimized to infer individual cell shapes in 4-cell mouse (top), 8-cell ascidian (middle), and 16-cell worm (bottom) embryos, using 3D fluorescence images acquired respectively with single-view, dual-view light-sheet, and point-scanning microscopes (see also: Supplemental videos 5–9). Mouse and ascidian data - courtesy of K. Cavanaugh and D. Gonzalez-Suarez.}
  \label{fig:10}
\end{figure*}

\subsection{Benchmarking against an active mesh method}
Traditional active mesh methods were primarily developed for segmenting bright volumetric regions~\cite{chanvese, dufour20103}, making them less suited for structures characterized by narrow intensity bands. We benchmark our approach against DM3D (Deforming Mesh 3D)~\cite{smith2023active}, a recent active mesh algorithm designed to detect cell interfaces in fluorescence microscopy images. Using the authors' dataset of 15 mouse organoid 3D images with corresponding inferred meshes, we reapply \textit{deltaMic} to the original image volumes. We then compare the outputs of \textit{DM3D} and \textit{deltaMic} using mean squared error (MSE), structural similarity index (SSIM)~\cite{ssim}, and peak signal-to-noise ratio (PSNR)~\cite{psnr}, computed between the reconstructed shapes rendered as synthetic images with our inferred PSF. Our method consistently outperforms DM3D across all metrics (Figure~\ref{fig:comparisonDM3D}).

\subsection{Shape inference from embryo microscopy}\label{celegans_ex}
We applied our method to infer cell shapes in early-stage embryos from different species and 3D fluorescence microscopy modalities: mouse (single-view light-sheet), ascidian (dual-view light-sheet), and worm (point-scanning confocal)~\cite{cao2020establishment}. Cell clusters were represented as non-manifold multimaterial meshes \cite{brakke1992surface,maitre2016asymmetric,firmin2024mechanics}, and we used a multimaterial mesh-based surface tracking method \cite{Multitracker} to handle remeshing, collision detection, and topological transitions. As an initial guess, we generated a foam-like multimaterial mesh with the correct number of cells. The resulting cell clusters, ranging between 1000–10000 vertices, were optimized to fit the microscopy data. We jointly optimized the PSF and mesh following the strategy in Section~\ref{strategies}. \textit{deltaMic} accurately reconstructed cell shapes in 4-, 8-, and 16-cell embryos, converging within 1k and 10k optimization steps, yielding rendered images that closely matched the original microscopy data (Figure~\ref{fig:10}).

\section{Conclusion}
We introduced \textit{deltaMic}, a first-of-its-kind differentiable 3D microscopy image creator that renders realistic 3D confocal fluorescence images from a surface mesh and a parametrized PSF. Our GPU-parallelized mesh FT implementation enables efficient forward and backward computations, further optimized with a spectral narrow-band method. \textit{deltaMic} is relatively robust to random noise in the image and mesh initialization, accurately recovering fine details of complex shapes from artificial and real 3D microscopy images thanks to its ability to fine-tune the shape loss function via the PSF. Unlike traditional active mesh methods, it does not require fine-tuned shape regularization terms to achieve smooth results. However, knowledge-based priors can be incorporated to constrain cell shapes, opening the avenue for direct implementation of inverse mechanical problems from microscopy data \cite{ichbiah2023embryo} through integration with a differentiable physical simulator.

Our approach opens numerous avenues for both practical applications and theoretical extensions. As an active contour method, it naturally lends itself to biological shape tracking in time-lapse microscopy. Incorporating more photorealistic PSF models tailored to specific microscopy modalities~\cite{Hanser:03,Hanser:04,Zernikereview,aguet2009super} could enable blind deconvolution of microscopy images with integrated shape priors~\cite{levin2009understanding,debarnot2024deep}. But extending our framework to handle space-variant PSFs~\cite{escande2015sparse,debarnot2021deepblur}, which violate translational invariance, may require significant theoretical and computational developments. In terms of geometry, the method could be adapted to handle 1D structures—such as cytoskeletal filaments, whose segmentation remains a major challe~\cite{ozdemir2021automated}—or extended to fully 3D shapes like nuclei or condensates, paving the way for multidimensional active mesh algorithms. On the implementation side, additional acceleration strategies could help mitigate the computational demands of large 3D volumes. Finally, thanks to its full differentiability, our method can be readily integrated as a module within larger (deep-)learning pipelines~\cite{DBLP:conf/cvpr/KatoUH18,DBLP:journals/corr/abs-2003-08934}.

\section{Acknowledgments}
SI, AS, FB and HT are supported by the CNRS and Collège de France. HT received funding from the European Research Council (ERC) under the European Union’s Horizon 2020 research and innovation programme (Grant agreement No. 949267) and from the French Agence Nationale de la Recherche (Grants ANR-22-CE13-0036 and ANR-23-CE30-0013). SI fellowship was funded by Ecole Polytechnique (AMX grant). We thank Baptiste Nicolet for insightful discussions, H. Borja da Rocha and K. Crane for sharing meshes, K. Cavanaugh and D. Gonzalez Suarez for sharing fluorescence microscopy images of mouse and ascidian embryos.

{\small
\bibliographystyle{ieeenat_fullname}
\bibliography{final}
}

\end{document}